\documentstyle[12pt]{article}
\makeatletter
\def\today{10 June 1996}
\advance\textheight by \topskip
\@ifundefined{selectfont}{
  \def\mathrm#1{{\rm #1}}
}{
  \let\oldmathrm=\mathrm
  \def\mathrm#1{{\oldmathrm{#1}}}
}
\@ifundefined{reset@font}{\let\reset@font\empty}{}
\def\bpsi{\mathord{\hbox{\boldmath$\psi$}}}
\def\bphi{\mathord{\hbox{\boldmath$\phi$}}}

\def\bxi{\mathord{\hbox{\boldmath$\xi$}}}
\def\bnu{\mathord{\hbox{\boldmath$\nu$}}}
\def\bbbone {{\mathchoice {\rm 1\mskip-4mu l} {\rm 1\mskip-4mu l}
{\rm 1\mskip-4.5mu l} {\rm 1\mskip-5mu l}}}
\def\thebibliography#1{\paragraph{References\@mkboth
 {REFERENCES}{REFERENCES}}\list
 {[\arabic{enumi}]}{\settowidth\labelwidth{[#1]}\leftmargin\labelwidth
 \advance\leftmargin\labelsep \itemsep=0pt
 \usecounter{enumi}}
 \def\newblock{\hskip .11em plus .33em minus .07em}
 \sloppy\clubpenalty4000\widowpenalty4000
 \sfcode`\.=1000\relax\small}
\makeatother
\def\vac{|0\rangle}
\def\id{{\bf 1}}
\begin{document}
\bibliographystyle{paper2}

\thispagestyle{empty}
\def\thefootnote{\fnsymbol{footnote}}
\begin{flushright}
  hep-th/9606050\\
DAMTP 96-54
\end{flushright}
\vskip 2.5em
\begin{center}\LARGE
  A Rational Logarithmic Conformal Field Theory
\end{center}\vskip 2em
\begin{center}\large
  Matthias R. Gaberdiel%
  \footnote{Email: {\tt M.R.Gaberdiel@damtp.cam.ac.uk}}%
  \footnote{Address from 1st September 1996: Department of
    Physics, Harvard University, Cambridge MA, 02138, USA} 
  and
  Horst G. Kausch%
  \footnote{Email: {\tt H.G.Kausch@damtp.cam.ac.uk}}%
  \footnote{Address from 9th September 1996: Department of
    Mathematics, King's College London, Strand, London WC2R 2LS, U.K.} 
\end{center}
\begin{center}\it
Department of Applied Mathematics and Theoretical
Physics, \\
University of Cambridge, Silver Street, \\
Cambridge CB3 9EW, U.K.
\end{center}
\vskip 1em
\begin{center}
  \today
\end{center}
\vskip 1em
\begin{abstract}
  We analyse the fusion of representations of the triplet algebra, the
  maximally extended symmetry algebra of the Virasoro algebra at
  $c=-2$. It is shown that there exists a finite number of
  representations which are closed under fusion. These include all
  irreducible representations, but also some reducible representations
  which appear as indecomposable components in fusion products.
\end{abstract}

\setcounter{footnote}{0}
\def\thefootnote{\arabic{footnote}}

\paragraph{1. Introduction.}

Recently a class of conformal field theories whose correlation
functions show logarithmic behaviour has attracted some
attention. These theories include for example the WZNW model on the
supergroup $GL(1,1)$ \cite{RSal92}, the $c=-2$ model \cite{Gur93},
gravitationally dressed conformal field theories \cite{BKog95} and
some critical disordered models \cite{CKT95,MSer96}. They are believed
to be important for the description of certain statistical models, in
particular in the theory of (multi)critical polymers
\cite{Sal92,Flohr95,Kau95} and percolation \cite{Watts96}. There have
also been suggestions that some of these logarithmic operators might
correspond to normalisable zero modes for string backgrounds
\cite{KMa95}.

These theories are also interesting from a more theoretical point of
view. They give rise to novel features in the representation theory of
the Virasoro algebra as the appearance of logarithms can be traced
back to the property of fusion products \cite{Gab93} to be not
completely reducible \cite{Gur93,Nahm94}. So far, all known examples
involved an infinite set of representations; in this letter, we
analyse a ``rational'' logarithmic conformal field theory, {\it i.e.}
a logarithmic conformal field theory for which a finite set of
representations closes under fusion.\footnote{ We should stress that
the notion of rationality we use here does not conform with the
mathematical literature: there rationality also entails that every
representation is completely decomposable into irreducibles, which
certainly is not true here.} Related issues have recently also
been discussed by Flohr \cite{Flohr95,Flohr96} and in
\cite{Kau95,GKau96,Roh96}.  
\smallskip

In \cite{GKau96} we introduced an algorithm to study the fusion
products of quasirational representations. We used this algorithm to
analyse the fusion of certain representations of the Virasoro $(1,q)$
models (the first of which, the $(1,2)$ model, has $c=-2$), and we
showed that there exists an infinite set of quasirational
representations which is closed under fusion. These representations
include irreducible representations, but also some reducible
representations which appear as indecomposable components in the
fusion products of irreducibles.

It has been known for some time \cite{Kausch91}, that the Virasoro
algebra, the chiral algebra of the Virasoro $(1,q)$ model, can be
extended by a triplet of weight $2q-1$ fields $W^i(z)$ to form the
{\it triplet algebra}. In this letter we shall mainly consider the
simplest case, where $q=2$, although much of what we say should also
hold in general. The infinitely many quasirational irreducible
representations of the Virasoro algebra organise themselves into
finitely many irreducible representations of this triplet algebra
\cite{Kau95}. Furthermore, as was argued in \cite{EHHu93}, the triplet
algebra has only finitely many irreducible representations. It is then
natural to ask whether there exists a finite number of
(indecomposable) representations of the triplet algebra which is
closed under fusion, and this is indeed what will turn out to be true.
\smallskip

It should be mentioned that Flohr \cite{Flohr96} has recently analysed
the modular properties of these models. He proposed a partition
function for the theory which is constructed from a certain finite set
of characters (including the characters of the irreducible
representations) which is invariant under the modular group. He then
applied the Verlinde formula to obtain ``fusion rule coefficients''
for corresponding representations, some of which turned out to be
negative.

Apart from the fact that the meaning of the negative fusion rule
coefficients is rather obscure, it is not clear from his point of view
which of his characters actually correspond to representations of the
triplet algebra, and if so, which structure they have. In particular,
different sets of possible (inequivalent) fusion rules can be found
which correspond to different choices for the basis of the
characters. It is also not clear in which way the feature of the
fusion products to be not completely reducible is taken into account. 

In our approach we can deduce the characters corresponding to the
finitely many representations which are closed under fusion. It turns
out that they are invariant under the modular group, but that a naive
application of Verlinde's formula does not make sense. However, there
exists some modified procedure (see case III in \cite{Flohr96}), by
means of which our results can be reproduced from character
considerations.

\paragraph{2. The triplet algebra and its irreducible representations.}

Let us recall the structure of the triplet algebra (at $c=-2$). It is
generated by the Virasoro modes $L_n$, and the modes of a triplet of
weight 3 fields $W^i_n$. The commutation relations are 
\begin{eqnarray}
  {}[ L_m, L_n ] &=& (m-n)L_{m+n} - \frac16 m(m^2-1) \delta_{m+n}, 
  \nonumber \\
  {}[ L_m, W^a_n ] &=& (2m-n) W^a_{m+n}, 
  \nonumber \\
  {}[ W^a_m, W^b_n ] &=& g^{ab} \biggl( 
  2(m-n) \Lambda_{m+n} 
  +\frac{1}{20} (m-n)(2m^2+2n^2-mn-8) L_{m+n} 
  \nonumber\\&&\qquad
  -\frac{1}{120} m(m^2-1)(m^2-4)\delta_{m+n}
  \biggr) 
  \nonumber \\&&
  + f^{ab}_c \left( \frac{5}{14} (2m^2+2n^2-3mn-4)W^c_{m+n} 
    + \frac{12}{5} V^c_{m+n} \right) \,, \nonumber
  \end{eqnarray}
where $\Lambda = \mathopen:L^2\mathclose: - 3/10\, \partial^2L$ and 
$V^a = \mathopen:LW^a\mathclose: - 3/14\, \partial^2W^a$ are
quasiprimary normal ordered fields. $g^{ab}$ and $f^{ab}_c$ are the
metric and structure constants of $su(2)$. In an orthonormal basis we
have $g^{ab} = \delta^{ab}, f^{ab}_c = i\epsilon^{abc}$. 

The triplet algebra (at $c=-2$) is only associative, because certain
states in the vacuum representation (which would generically violate
associativity) are null. The relevant null vectors are
\begin{eqnarray} 
  N^a &=& 
  \left(2 L_{-3} W^a_{-3} -\frac43 L_{-2} W^a_{-4} + W^a_{-6}\right)
  \vac, 
\label{eq:nulllw}
\\
  N^{ab} &=&
  W^a_{-3} W^b_{-3} \vac - 
  g^{ab} \left( \frac89 L_{-2}^3 + \frac{19}{36} L_{-3}^2 +
  \frac{14}{9} L_{-4}L_{-2} - \frac{16}{9} L_{-6} \right)\vac 
  \nonumber\\*&&
  - f^{ab}_c\left( -2 L_{-2} W^c_{-4} + \frac54 W^c_{-6}
  \right) \vac \,.
\label{eq:nullww}
\end{eqnarray}
We shall only be interested in representations which respect these
relations, and for which the spectrum of $L_0$ is bounded from below.
Then, the zero modes of the null-states have to vanish on the ground
states, and this will restrict the allowed representations
\cite{FNO92}. Evaluating the constraint coming from (\ref{eq:nullww}),
we find 
\begin{equation}
  \left(W^a_0 W^b_0 - g^{ab} \frac19 L_0^2 (8L_0 + 1) - 
  f^{ab}_c \frac15 (6L_0-1) W^c_0 \right) \psi = 0 \,,
\label{eq:wwzero}
\end{equation}
where $\psi$ is any ground state, while the relation coming from
the zero mode of (\ref{eq:nulllw}) is satisfied identically.
Furthermore, the constraint from $W^a_1 N^{bc}_{-1}$, together with
(\ref{eq:wwzero}) implies that $W^a_0 (8 L_0 - 3) (L_0 -1) \psi = 0$. 
Multiplying with $W_0^a$ and using (\ref{eq:wwzero}) again, this implies that
\begin{equation}
\label{eq:heigen}
0  = L_0^2 (8 L_0 + 1) (8 L_0 - 3) (L_0 - 1) \psi\,.
\end{equation}

For irreducible representations, $L_0$ has to take a fixed value $h$
on the ground states, and (\ref{eq:heigen}) then implies that $h$
has to be either $h=0, -1/8, 3/8$ or $h=1$. It also follows from 
(\ref{eq:wwzero}) that 
\begin{displaymath}
  {}[ W^a_0, W^b_0 ] = \frac25 (6h-1) f^{ab}_c W^c_0 \,,
\end{displaymath}
which is a rescaled version of the $su(2)$ algebra. After rescaling, 
the irreducible representations of these zero modes can then be
labelled by $j$ and $m$, where $j(j+1)$ is the  eigenvalue of the
Casimir operator $\sum_a (W^a_0)^2$, and $m$ is the eigenvalue of
$W^3_0$. Because of (\ref{eq:wwzero}), $W^a_0 W^a_0 = W^b_0
W^b_0$ on the ground states, and thus $j(j+1)=3m^2$. This can
only be satisfied for $j=0, 1/2$, and this restricts the allowed
representations to 
\begin{itemize}
\item the singlet representations, $j=0$, at $h=0, -1/8$\,;
\item the doublet representations, $j=1/2$, at $h=1, 3/8$\,,
\end{itemize}
as was remarked in \cite{EHHu93}. 

These four representations have been constructed from a free field
realisation in \cite{Kau95}. We shall now briefly discuss their
structure, paying particular attention to the special subspace, one of
the central concepts in the analysis of fusion \cite{Nahm94,GKau96}.

\paragraph{2a. The singlet representation at $h=0$:}

This is the vacuum representation, ${\cal V}_0$. We denote the vacuum
state by $\Omega$, and as usual, we have
\begin{equation}
\label{eq:vacuum}
\begin{array}{rcl}
  L_{-n} \Omega &=& 0, \qquad\mbox{for $n\leq1$}\,, \\
  W^a_{-n} \Omega &=& 0, \qquad\mbox{for $n\leq2$}\,.
\end{array}
\end{equation}
This implies that the special subspace is trivial, 
${\cal V}_0^{\mathrm{s}} = \langle\Omega\rangle$. Further relations in
the vacuum representation are given by (\ref{eq:nulllw}) and
(\ref{eq:nullww}).   

\paragraph{2b. The singlet representation at $h=-1/8$:}

We denote the ground state of ${\cal V}_{-1/8}$ by $\mu$. The
constraints coming from the negative modes of the vacuum null vectors
(\ref{eq:nulllw}) and (\ref{eq:nullww}) yield 
$W^a_0 \mu=W^a_{-1} \mu = 0$, $L_{-1}^2 \mu= 1/2\, L_{-2} \mu$,
$L_{-1} W^a_{-2} \mu = 3/4\, W^a_{-3} \mu$, and  
\begin{equation}
\label{eq:singlet}
  W^a_{-2} W^b_{-2} \mu = g^{ab} \left( L_{-2}^2 + \frac52
          L_{-3}L_{-1} - \frac78 L_{-4} \right) \mu 
 +f^{ab}_c \left( 2 L_{-2}W^c_{-2} + \frac14 W^c_{-4} \right) \mu\,.
\end{equation}
It follows that the special subspace is five-dimensional, and that a
basis can be chosen to consist of 
$$
  {\cal V}_{-1/8}^{\mathrm{s}} = \langle\mu, L_{-1}\mu,
  W^a_{-2}\mu\rangle\,. 
$$

\paragraph{2c. The doublet representation at $h=1$:} 

We denote the two ground states of ${\cal V}_{1}$ by $\psi^\pm$. To
specify the action of the $W_0$ modes on $\psi^\pm$, let us introduce
a Cartan-Weyl basis $W^0, W^\pm$ for $su(2)$, so that the
non-vanishing structure constants are $f^{0\pm}_{\pm} =
-f^{\pm0}_{\pm} = \pm1$ and $f^{+-} = -f^{-+} = 2$. We also define  
a metric by $g^{00} = 1$, $g^{+-} = g^{-+} = 2$. On the states
$\psi^\pm$ of the spin $1/2$ representation of $su(2)$, we then define
the representation matrices $t^a$ by
\begin{displaymath}
  t^0 \psi^\pm = \pm\frac12 \psi^\pm\,, \qquad
  t^\pm \psi^\pm = 0\,, \qquad
  t^\pm \psi^\mp = \psi^\pm\,.
\end{displaymath}
In this notation, the constraints coming from the null vectors
(\ref{eq:nulllw}) and (\ref{eq:nullww}) yield 
\begin{equation}
\label{eq:doublet1}
\begin{array}{rclrcl}
  W^a_{0} \bpsi &=& 2t^a\bpsi\,, \hspace*{2cm} 
     & W^a_{-1} \bpsi &=& 3L_{-1}t^a\bpsi\,, \\ 
  W^a_{-2} \bpsi &=& 4L_{-2}t^a\bpsi\,, &  L_{-1}^2\bpsi &=& 2L_{-2}\bpsi\,.
\end{array}
\end{equation}
It follows that the special subspace is four-dimensional, and that a
basis can be chosen to consist of  
$$
  {\cal V}_{1}^{\mathrm{s}} = \langle\psi^\pm, L_{-1}\psi^\pm\rangle. 
$$

\paragraph{2d. The doublet representation at $h=3/8$:}

Finally, we denote the two ground states of ${\cal V}_{3/8}$ by
$\nu^\pm$. The constraints give in this case 
$W^a_{0} \bnu = 1/2 t^a\bnu$, $W^a_{-1} \bnu = 2L_{-1}t^a\bnu$, and
{\renewcommand{\arraystretch}{2}
\begin{equation}
\label{eq:doublet38}
\begin{array}{rcl}
  W^a_{-2} \bnu &=& {\displaystyle
          \left(\frac32L_{-2}+L_{-1}^2\right)t^a\bnu\,,} \\
  L_{-1}^4\bnu &=& {\displaystyle 
     \left( 5L_{-2}L_{-1}^2 - \frac94L_{-2}^2 -
  L_{-3}L_{-1} + \frac32L_{-4}\right)\bnu\,.}
\end{array}
\end{equation}
}%
It follows that the special subspace is eight-dimensional, and that a
basis can be chosen to be
$$
  {\cal V}_{3/8}^{\mathrm{s}} = \langle\nu^\pm, L_{-1}\nu^\pm, L_{-1}^2\nu^\pm,
  L_{-1}^3\nu^\pm\rangle\,.  
$$

Any highest weight representations of the triplet algebra of conformal
weight $h=-1/8, 1$ or $3/8$ is automatically irreducible, and
therefore isomorphic to ${\cal V}_{-1/8}, {\cal V}_1$ and ${\cal V}_{3/8}$,
respectively. This follows from the fact that a reducible
representation would have to contain a subrepresentation whose highest
weight would be a descendant singular vector. However, the above
analysis implies that such singular vectors do not exist. 

On the other hand, there exist reducible highest weight
representations with $h=0$. In particular, the highest weight
representation ${\cal M}_0$, generated from a ground state $\omega$ of
conformal weight $h=0$ and $j=0$, which does not satisfy
(\ref{eq:vacuum}), but only the constraints coming from
(\ref{eq:nulllw}) and (\ref{eq:nullww}) is reducible. It contains 
two subrepresentations of type ${\cal V}_1$, whose  ground
states are given as 
\begin{displaymath}
  \begin{array}{rcl@{\qquad}rcl}
    \psi_1^+ &=& W^+_{-1} \omega\,, & 
    \psi_2^+ &=& \left(W^0_{-1} + \frac12 L_{-1}\right) \omega\,, \\
    \psi_1^- &=& \left(-W^0_{-1} + \frac12 L_{-1}\right) \omega\,, &
    \psi_2^- &=& W^-_{-1} \omega\,.
  \end{array}
\end{displaymath}

\paragraph{3. Fusion and indecomposable representations.}

To calculate the fusion of these representations we use the same
algorithm as in \cite{GKau96} to which we also refer for more
details. Let us illustrate the method by explaining a simple case
explicitly, the fusion of ${\cal V}_{-1/8}\times{\cal V}_{1}$. On
general grounds, as was shown in \cite{Nahm94}, we know that the space
of lowest energy states in the fusion product is contained in 
${\cal V}_{-1/8}^{\mathrm{s}}\otimes{\cal V}_{1}^0$. However, this
ten-dimensional space is too large since we have
\begin{eqnarray*}
  \Delta(W^a_{-2}) (\mu\otimes\bpsi) - 
  4\Delta(L_{-2}) (\mu\otimes t^a\bpsi)
  &=& W^a_{-2}(\mu\otimes\bpsi) - 4(L_{-1}\mu\otimes t^a\bpsi)
\nonumber\\*&&
  - \frac12 (\mu\otimes t^a\bpsi)\,, \\
  \Delta(W^a_{-1}) (\mu\otimes\bpsi) - 
  3\Delta(L_{-1}) (\mu\otimes t^a\bpsi)
  &=& W^a_{-2}(\mu\otimes\bpsi) - 3(L_{-1}\mu\otimes t^a\bpsi)\,.
\end{eqnarray*}
To lowest level in the fusion product we can thus replace
\begin{displaymath}
  W^a_{-2}\mu\otimes\bpsi \mapsto -\frac32 \mu\otimes t^a\bpsi\,, 
  \qquad L_{-1}\mu\otimes\bpsi \mapsto -\frac12 \mu\otimes\bpsi\,,
\end{displaymath}
and choose representatives $\mu\otimes\psi^\pm$ for the
two-dimensional ground space $({\cal V}_{-1/8}\otimes 
{\cal V}_{1})_{\mathrm{f}}^0$. The action of $L_0$ and $W^a_0$ on this
ground space is easily determined as
\begin{displaymath}
  \Delta(L_0) = \frac38 \id, \qquad
  \Delta(W^a_0) = \frac12 t^a\,,
\end{displaymath}
and the fusion product is thus a highest weight representation of type
$h=3/8, j=1/2$. As was argued before, this representation is
automatically irreducible, and therefore it is not necessary to
analyse the states at higher level. We have thus shown that\footnote{
We should mention that our calculations only imply upper bounds for
the fusion rules. This is however sufficient to prove rationality.}
\begin{equation}
  ({\cal V}_{-1/8}\otimes{\cal V}_{1})_{\mathrm{f}} 
          = {\cal V}_{3/8}\,.
\end{equation}
In the same way we can show that the fusion with ${\cal V}_0$ is
trivial, and that 
\begin{equation}
  ({\cal V}_{3/8}\otimes{\cal V}_{1})_{\mathrm{f}} = {\cal V}_{-1/8}\,.
\end{equation}

The remaining fusion products involve potentially reducible
representations. A calculation to lowest level is then not sufficient,
and we analysed the product spaces up to level two. Before describing
the results in more detail, let us comment briefly on the question
whether the algorithm terminates. It is clear from
(\ref{eq:singlet}\,, \ref{eq:doublet1}\,, \ref{eq:doublet38}) that we
can replace successively the $W$-modes by $L$-modes, and that the
special subspace relations for the $L$-modes close among
themselves. It is then clear, by the same argument as in
\cite{GKau96}, that the algorithm always stops.

In the remaining cases we found 
\begin{equation}
\begin{array}{rcl}
  \left({\cal V}_{1}\otimes{\cal V}_{1}\right)_{\mathrm{f}} &=& 
    {\cal V}_0\,, \\
  \left({\cal V}_{-1/8}\otimes{\cal V}_{-1/8}\right)_{\mathrm{f}} &=& 
    {\cal R}_{0}\,, \\
  \left({\cal V}_{-1/8}\otimes{\cal V}_{3/8}\right)_{\mathrm{f}} &=& 
    {\cal R}_{1}\,, \\
  \left({\cal V}_{3/8}\otimes{\cal V}_{3/8}\right)_{\mathrm{f}} &=& 
    {\cal R}_{0}\,,
\end{array}
\end{equation}
where ${\cal R}_{0}, {\cal R}_{1}$ are generalised highest weight
representations whose structure can (schematically) be described by
the following diagram:
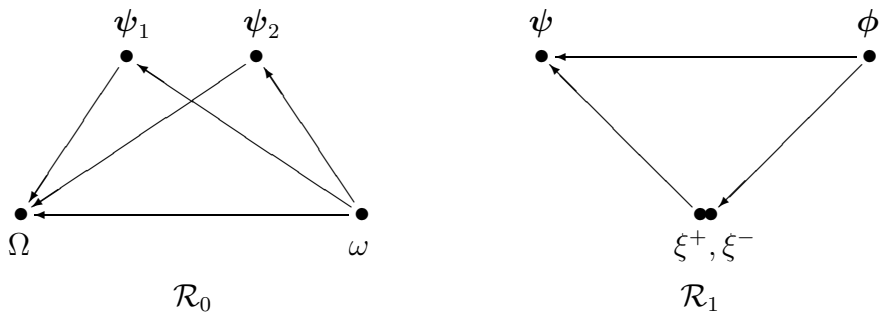
\begin{displaymath}
  \begin{array}{c@{\qquad\qquad}c}
    \begin{picture}(150,120)(-10,-20)
      \put(0,0){\vbox to 0pt
        {\vss\hbox to 0pt{\hss$\bullet$\hss}\vss}}
      \put(129,0){\vbox to 0pt
        {\vss\hbox to 0pt{\hss$\bullet$\hss}\vss}}
      \put(40,60){\vbox to 0pt
        {\vss\hbox to 0pt{\hss$\bullet$\hss}\vss}}
      \put(89,60){\vbox to 0pt
        {\vss\hbox to 0pt{\hss$\bullet$\hss}\vss}}
      \put(37,56){\vector(-2,-3){34}}
      \put(85,57){\vector(-3,-2){81}}
      \put(125,3){\vector(-3,2){81}}
      \put(126,5){\vector(-2,3){34}}
      \put(124,0){\vector(-1,0){119}}
      \put(-5,-15){$\Omega$}
      \put(124,-15){$\omega$}
      \put(35,70){$\bpsi_1$}
      \put(84,70){$\bpsi_2$}
    \end{picture}
    &
    \begin{picture}(140,120)(-10,-20)
      \put(0,60){\vbox to 0pt
        {\vss\hbox to 0pt{\hss$\bullet$\hss}\vss}}
      \put(124,60){\vbox to 0pt
        {\vss\hbox to 0pt{\hss$\bullet$\hss}\vss}}
      \put(60,0){\vbox to 0pt
        {\vss\hbox to 0pt{\hss$\bullet$\hss}\vss}}
      \put(64,0){\vbox to 0pt
        {\vss\hbox to 0pt{\hss$\bullet$\hss}\vss}}
      \put(119,60){\vector(-1,0){114}}
      \put(57,3){\vector(-1,1){54}}
      \put(121,57){\vector(-1,-1){54}}
      \put(-5,70){$\bpsi$}
      \put(119,70){$\bphi$}
      \put(50,-15){$\xi^+,\xi^-$}
    \end{picture}
    \\
    {\cal R}_0 & {\cal R}_1
  \end{array}
\end{displaymath}
Here each vertex represents an irreducible representation, 
${\cal V}_0$ in the bottom row and ${\cal V}_1$ in the top row. 
An arrow $A\longrightarrow B$ indicates that the vertex $B$ is in the
image of $A$ under the action of the triplet algebra.

\paragraph{3a. The representation ${\cal R}_{0}$:}

This representation is generated from a cyclic vector $\omega$ of
$h=0$ forming a Jordan block with $\Omega$. 
It
is an extension of ${\cal V}_0$ by ${\cal
  M}_0$. The defining
relations are 
\begin{eqnarray}
  L_0 \omega &=& \Omega\,, \nonumber \\
  W^a_0 \omega &=& 0\,. \nonumber
\end{eqnarray}
This is the only possible generalised highest weight representation
which can be obtained from a two-dimensional Jordan block at $h=0$.  It
also follows from (\ref{eq:heigen}) that larger Jordan blocks at $h=0$
are incompatible with the vacuum null-relations.

\paragraph{3b. The representation ${\cal R}_{1}$:}

This representation is generated from a doublet $\phi^\pm$ of cyclic
states of weight $h=1$. It has two ground states
$\xi^\pm$ at $h=0$ and another doublet $\psi^\pm$ at $h=1$ forming
$L_0$ Jordan blocks with $\phi^\pm$. The defining relations are
\begin{displaymath}
  \begin{array}{rcl@{\qquad}rcl}
    L_0 \bpsi &=& \bpsi\,, &
    W^a_0 \bpsi &=& 2t^a \bpsi\,,  \\[\bigskipamount]
    L_0 \bxi &=& 0\,, &
    W^a_0 \bxi &=& 0\,, \\
    L_{-1} \bxi &=& \bpsi\,, &
    W^a_{-1} \bxi &=& t^a \bpsi\,, \\[\bigskipamount]
    L_1 \bphi &=& -\bxi\,, &
    W^a_1 \bphi &=& -t^a \bxi\,, \\
    L_0 \bphi &=& \bphi + \bpsi\,, &
    W^a_0 \bphi &=& 2t^a \bphi\,.
  \end{array}
\end{displaymath}
We stress that $\psi^\pm$ and $\phi^\pm$ form a Jordan block with
respect to $L_0$ but that they remain uncoupled with respect to
$W^a_0$. On higher levels there are also Jordan blocks for $W^a_0$. 

At first, it would seem that there might be inequivalent
representations of this structure, as were obtained for the Virasoro
case in \cite{GKau96} (see also \cite{Roh96}). However, in contrast
to the situation for the Virasoro algebra, the vacuum representation
of the triplet algebra contains null vectors, and the compatibility
with these constraints restricts this freedom. It turns out, that
there is no free parameter, and that the representation given above is
the only one of its type. 

The indecomposable representations ${\cal R}_0$ and ${\cal R}_1$ of
the triplet algebra decompose (with respect to the Virasoro
subalgebra) into the indecomposable representations 
${\cal R}_{2m-1,1}$ and ${\cal R}_{2m,1}$ of the Virasoro algebra.  
In particular, as a subrepresentation of ${\cal R}_0$ and ${\cal R}_1$, 
the characteristic parameters of these Virasoro representations
are fixed. We have determined these parameters, and found that they
agree in all cases (${\cal R}_{m,1}$ with $m\leq5$) with those
determined in \cite{GKau96}. The representations of the Virasoro
algebra appearing in fusion products at $c=-2$ are therefore precisely
those which come from the allowed representations of the triplet
algebra.  
\bigskip

Finally, we calculated the fusion products involving the indecomposable
representations. For the first four fusion products a calculation to
lowest level was again sufficient; the other cases were calculated 
to level two.
\begin{eqnarray}
  \left({\cal V}_{-1/8}\otimes{\cal R}_{0}\right)_{\mathrm{f}} &=& 
    2{\cal V}_{-1/8} \oplus 2{\cal V}_{3/8}\,, \nonumber \\
  \left({\cal V}_{-1/8}\otimes{\cal R}_{1}\right)_{\mathrm{f}} &=& 
    2{\cal V}_{-1/8} \oplus 2{\cal V}_{3/8}\,, \nonumber \\
  \left({\cal V}_{3/8}\otimes{\cal R}_{0}\right)_{\mathrm{f}} &=& 
    2{\cal V}_{-1/8} \oplus 2{\cal V}_{3/8}\,, \nonumber \\
  \left({\cal V}_{3/8}\otimes{\cal R}_{1}\right)_{\mathrm{f}} &=& 
    2{\cal V}_{-1/8} \oplus 2{\cal V}_{3/8}\,, \nonumber
\end{eqnarray}
\begin{eqnarray}
  \left({\cal V}_{1}\otimes{\cal R}_{0}\right)_{\mathrm{f}} &=& 
    {\cal R}_{1}\,, \nonumber \\
  \left({\cal V}_{1}\otimes{\cal R}_{1}\right)_{\mathrm{f}} &=& 
    {\cal R}_{0}\,, \nonumber \\[\bigskipamount]
  \left({\cal R}_{0}\otimes{\cal R}_{0}\right)_{\mathrm{f}} &=& 
    2{\cal R}_{0} \oplus 2{\cal R}_{1}\,, \nonumber \\
  \left({\cal R}_{0}\otimes{\cal R}_{1}\right)_{\mathrm{f}} &=& 
    2{\cal R}_{0} \oplus 2{\cal R}_{1}\,, \nonumber \\
  \left({\cal R}_{1}\otimes{\cal R}_{1}\right)_{\mathrm{f}} &=& 
    2{\cal R}_{0} \oplus 2{\cal R}_{1}\,. \nonumber
\end{eqnarray}
It follows from these results that the set of representations 
${\cal V}_{0}, {\cal V}_{1}, {\cal V}_{-1/8}, {\cal V}_{3/8}$ and 
${\cal R}_{0}, {\cal R}_{1}$, is closed under fusion. The triplet 
algebra at $c=-2$ defines therefore a rational logarithmic conformal
field theory.

It is natural to speculate that this conclusion will also hold for the
other $(1,q)$ models. We have checked this explicitly for the next
case, the triplet algebra at $c=-7$, and the results confirm this
conjecture.

\paragraph{4. Characters and modular transformations.}

The characters of the irreducible representations of the triplet
algebra have been calculated in \cite{Kau95} (see also
\cite{Flohr95}). From these, and the explicit description of the
indecomposable representations, we can derive the characters of all
the above representations. In more detail we have
\begin{eqnarray}
  \chi_{{\cal V}_0}(\tau) &=&
          \frac12\left(\eta(\tau)^{-1}\theta_{1,2}(\tau) +
              \eta(\tau)^2\right)\,, \nonumber \\ 
  \chi_{{\cal V}_1}(\tau) &=& 
          \frac12\left(\eta(\tau)^{-1}\theta_{1,2}(\tau) -
              \eta(\tau)^2\right)\,, \nonumber \\ 
  \chi_{{\cal V}_{-1/8}}(\tau) &=& \eta(\tau)^{-1}
\theta_{0,2}(\tau)\,, \nonumber  \\
  \chi_{{\cal V}_{3/8}}(\tau) &=& \eta(\tau)^{-1}
\theta_{2,2}(\tau)\,, \nonumber  \\
  \chi_{{\cal R}_0}(\tau) &=& 2\eta(\tau)^{-1} \theta_{1,2}(\tau)\,,
\nonumber \\
  \chi_{{\cal R}_1}(\tau) &=& 2\eta(\tau)^{-1}
\theta_{1,2}(\tau)\,. \nonumber  
\end{eqnarray}
It turns out that the space generated by the last four characters is
invariant under the action of the modular group, and that each has a
suitable transformation property under $S$, whereas the $S$
transformation of $\chi_{{\cal V}_0}(\tau)$ and 
$\chi_{{\cal V}_1}(\tau)$ involves coefficients which are themselves
functions of $\tau$, {\it i.e.}  
\begin{eqnarray*}
  \chi_{{\cal V}_0}(-1/\tau) &=& \frac14 \chi_{{\cal V}_{-1/8}}(\tau) - 
  \frac14 \chi_{{\cal V}_{3/8}}(\tau) - \frac{i\tau}{2}
         \eta(\tau)^2\,, 
  \\
  \chi_{{\cal V}_1}(-1/\tau) &=& \frac14 \chi_{{\cal V}_{-1/8}}(\tau) - 
  \frac14 \chi_{{\cal V}_{3/8}}(\tau) + \frac{i\tau}{2} \eta(\tau)^2\,.  
\end{eqnarray*}
There have been attempts to give an interpretation of this
transformation property in terms of generalised characters
\cite{Flohr95}.

The representations ${\cal R}_0$ and ${\cal R}_1$ contain the
irreducible subrepresentations ${\cal V}_0$ and ${\cal V}_1$,
respectively, and the set of representations ${\cal V}_{-1/8}, {\cal
  V}_{3/8}$ and ${\cal R}_{0}, {\cal R}_{1}$ is already closed under
fusion. This suggests that the fundamental building blocks of the
theory are these four representations ({\it c.f.} also \cite{Roh96}).
Two of its characters are the same, and so the definition of the
modular matrices is ambiguous. It turns out that there is a one
parameter freedom to define these matrices so that the relations of
the modular group, $S^4=\bbbone$ and $(ST)^3=S^2$ are
satisfied,\footnote{All different choices lead to equivalent
representations of the modular group.} and a unique solution for which
the charge conjugation matrix $S^2$ is a permutation matrix. In the
basis of $\chi_{{\cal R}_0}, \chi_{{\cal R}_1}, 
\chi_{{\cal V}_{-1/8}}, \chi_{{\cal V}_{3/8}}$ this solution is given
as  
$$
S = \left(\matrix{-\frac{1}{2} i & \frac{1}{2} i & \frac{1}{4} & 
-\frac{1}{4} \vspace*{0.2cm} \cr 
\frac{1}{2} i & - \frac{1}{2} i & \frac{1}{4} & 
-\frac{1}{4} \vspace*{0.2cm} \cr
1 & 1 & \frac{1}{2} & \frac{1}{2} \vspace*{0.2cm} \cr
-1 & -1 & \frac{1}{2} & \frac{1}{2}}\right) 
\hspace*{0.5cm}
T= \left(\matrix{e^{2\pi i / 12} & 0 & 0 & 0 \cr
0 & e^{2\pi i / 12} & 0 & 0 \cr
0 & 0 & e^{- \pi i / 12} & 0 \cr
0 & 0 & 0 & e^{11 \pi i / 12} }\right) \,.
$$
It is intriguing that a formal application of Verlinde's
formula leads to fusion rule coefficients which are positive integers. 
These do not reproduce the fusion rules we
have calculated. This is not surprising, as, for example, this set of 
representations does not contain the vacuum representation, {\it i.e.}
a representation which has trivial fusion rules. Even more
drastically, the fusion matrix corresponding to the representation 
${\cal V}_{-1/8}$ which, in the above basis, is given as
$$
N_{-1/8} = \left(\matrix{0 & 0 & 1 & 0 \cr 
0 & 0 & 0 & 1 \cr
2 & 2 & 0 & 0\cr
2 & 2 & 0 & 0} \right)\,,
$$ 
is not diagonalisable, and the same is true for the matrix
corresponding to ${\cal V}_{3/8}$. However, a slight modification of
Verlinde's observation still holds: the above $S$ matrix transforms
the fusion matrices into block diagonal form, where the blocks correspond
to the two one-dimensional and the one two-dimensional representation
of the fusion algebra.

\paragraph{Acknowledgements.}

We would like to thank Wolfgang Eholzer, Michael Flohr and G\'erard
Watts for useful discussions.

M.R.G. is supported by a Research Fellowship of Jesus College,
Cambridge, and H.G.K. by a Research Fellowship of Sidney Sussex
College, Cambridge. This work has also been supported in part by
PPARC. 

The computer calculations were performed on computers purchased on
EPSRC grant GR/J73322, using a MAPLE package written by H.G.K.

\end{document}